\documentstyle[epsfig]{mn}

\ifCUPmtlplainloaded \else
   \DeclareSymbolFont{UPM}{U}{eur}{m}{n}
   \SetSymbolFont{UPM}{bold}{U}{eur}{b}{n}
   \DeclareMathSymbol{\upi}{0}{UPM}{"19}
   \DeclareMathSymbol{\umu}{0}{UPM}{"16}
   \DeclareMathSymbol{\upartial}{0}{UPM}{"40}
\fi

\begin{document}

\title[Excess sub-mm emission from GRS 1915+105]{Excess sub-millimetre
emission from GRS 1915+105}

\author[R. N. Ogley et al.]
{R. N. Ogley,\thanks{E-mail:ogley@discovery.saclay.cea.fr}$^{1}$
S. J. Bell Burnell$^{2,3}$,
R. P. Fender$^4$,
G. G. Pooley$^5$\cr
and E. B. Waltman$^6$\\
$^1$Service d'Astrophysique, C.E.A. Saclay, Orme des Merisiers -
B\^{a}t 709, F-91191 Gif sur Yvette Cedex, France\\
$^2$Department of Physics, The Open University, Milton Keynes, MK7
6AA\\
$^3$Princeton University, Physics Department, Jadwin Hall, PO Box 708, Princeton, NJ 08544-0708, USA\\
$^4$Astronomical Institute `Anton Pannekoek' and Center for
High-Energy Astrophysics, University of Amsterdam,\\ 
Kruislaan 403, 1098 SJ Amsterdam, The Netherlands\\
$^5$MRAO, Cavendish Laboratory, University of Cambridge, CB3 0HE\\
$^6$Remote Sensing Division, Naval Research Laboratory, Code 7210,
Washington, DC 20375-5351, USA}

\maketitle
\begin{abstract}

We present the first detections of the black hole X-ray binary GRS
1915+105 at sub-millimetre wavelengths. We clearly detect the source
at 350 GHz on two epochs, with significant variability over the 24 hr
between epochs. Quasi-simultaneous radio monitoring indicates an
approximately flat spectrum from 2 -- 350 GHz, although there is
marginal evidence for a minimum in the spectrum between 15 -- 350
GHz. The flat spectrum and correlated variability imply that the
sub-mm emission arises from the same synchrotron source as the radio
emission. This source is likely to be a quasi-steady partially
self-absorbed jet, in which case these sub-mm observations probe 
significantly closer to the base of the jet than do radio observations
and may be used in future as a valuable diagnostic of the disc:jet
connection in this source.

\end{abstract}


\section{Introduction}

Since its discovery in 1992, simultaneously with the SIGMA instrument
on the {\it Granat} satellite and the BATSE instrument on the {\it
CGRO} satellite (Castro-Tirado, Brandt, Lund, 1992; Harmon, Paciesas,
Fishman, 1992), GRS 1915+105 has been one of the most extensively
studied sources of recent times.  The popularity of this source is due
to its highly complex and variable nature at all wavelengths from
gamma rays to radio.  The many unusual features include relativistic
jets (Mirabel \& Rodr\'{\i}guez 1994; Fender et al.\ 1999), X-ray
Quasi-periodic Oscillations (QPOs) (Morgan \& Remillard 1996) and
accretion instabilities resulting in jet formation (Pooley \& Fender
1997; Eikenberry et al. 1998; Mirabel et al. 1998).

The system exhibits a wide variety of radio behaviour on timescales
from minutes to weeks (Foster et al. 1996; Pooley \& Fender 1997).
High time resolution observations using the Ryle Telescope (RT) and
more continuous coverage on hourly time-scales, revealed new aspects
of the radio emission (Pooley \& Fender 1997 hereafter PF97).  Both
20--40 min period QPOs associated with soft X-ray variations (first
reported in the radio by Pooley (1995,1996)) and a change of QPO
period, were clearly visible during some individual observations.

The infrared variability from this source is similar. Following
observed 1 and 2 mag variations in the {\it J}, {\it H} and {\it
K}-band fluxes (Castro-Tirado et al.\ 1993; Mirabel et al.\ 1994;
Chaty et al.\ 1996), Fender et al.\ (1997) reported rapid infrared
flares which had amplitudes, rise, decay and recurrence time-scales
strikingly similar to those of the radio flares observed with the RT 8
hours later, suggesting infrared synchrotron emission.  Further RT
observations taken simultaneously with the William Herschel Telescope
(WHT) showed 26-min oscillations in both the radio and infrared
emission (Fender \& Pooley 1998).  From these observations, the
authors showed that the radio variations were delayed by 33 (or
perhaps 59) minutes relative to the infrared.

Is there a correlation between the radio or infrared emission and the
X-rays?  There is clearly some correlation between X-ray states and
radio emission ({\it CGRO} Burst And Transient Source Experiment
(BATSE) correlation reported by Foster et al.\ 1996, {\it Rossi X-ray
Timing Explorer} Proportional Counter Array ({\it RXTE} PCA)
correlation reported by PF97), and PF97 were also the first to observe
frequency-dependent delays in the radio emission.  Quasi-periodic
flaring in the infrared, reported by Fender \& Pooley (1998), was also
observed by Eikenberry et al.\ (1998), who had simultaneous
observations with {\it RXTE} PCA.  The source of these flares is
consistent with infrared emitting plasmons from {\it ejected disc
material}, following rapid X-ray variability and an X-ray dip.  The
confirmation of such a hypothesis was presented by Mirabel et al.\
(1998) with simultaneous {\it RXTE} PCA, United Kingdom Infrared
Telescope (UKIRT) and Very Large Array (VLA) data.  Several X-ray dips
and subsequent infrared and radio flares were observed.

\section{Photometry}

We obtained sub-mm data over two days on MJD 51026 and 51027, when GRS
1915+105 was moderately active, with periods of quiescence
interspersed.  Co-ordinated with this were quasi-simultaneous
observations at radio wavelengths.  The radio data were taken using the
Green Bank Interferometer (GBI) at 2.25 and 8.3 GHz, and the RT at
15 GHz.  Sub-mm data at 350 GHz was taken using the James Clark
Maxwell Telescope (JCMT) using the SCUBA array, and X-ray data were
obtained at 2--12 keV from the {\it RXTE} ASM instrument.

The GBI monitored GRS 1915+105 simultaneously at 2.25 and 8.3 GHz with
integrations 10--15 minutes in length several times daily.  The
calibration procedure is reported in detail in Waltman et al.\ (1994).
The RT is an 8-element east--west array operating at 15-GHz; details
of the operation can be found in PF97.  The JCMT Sub-mm Common User
Bolometer Array (SCUBA) observed GRS 1915+105 over two nights on MJD
51026 and 51027 (1998 Aug 01 and 02) in photometric mode at 450 and
850 $\umu$m wavelengths -- for a description of the instrument see
Holland et al.\ (1999).  Regular skydips were taken throughout the
observations, and after focusing, aligning and pointing the array,
flux calibration was performed using either Uranus or G45.1
($\rm{RA}(1950)=19^{\rm{h}}~11^{\rm{m}}$,
$\rm{Dec.}~(1950)=10^{\circ}~45'$).  A total of 15 observations of GRS
1915+105 were made over the two nights, each observation included 50
integrations of 10 s (except for the observation at 51026.574
which included 26 integrations).

Fig.\ \ref{1026,1027_photometry} shows photometry over two nights: MJD
51026 and 51027.  In the figure, the top panel shows the X-ray flux
from the {\it RXTE} ASM instrument, whereas the bottom panel shows the
radio and sub-mm flux.  Table \ref{weighted_average} presents the flux
range during observations, a typical observation error for each
photometric point, a weighted mean, a projected flux (detailed below),
and errors on these fluxes.
\begin{figure*}
\epsfig{file=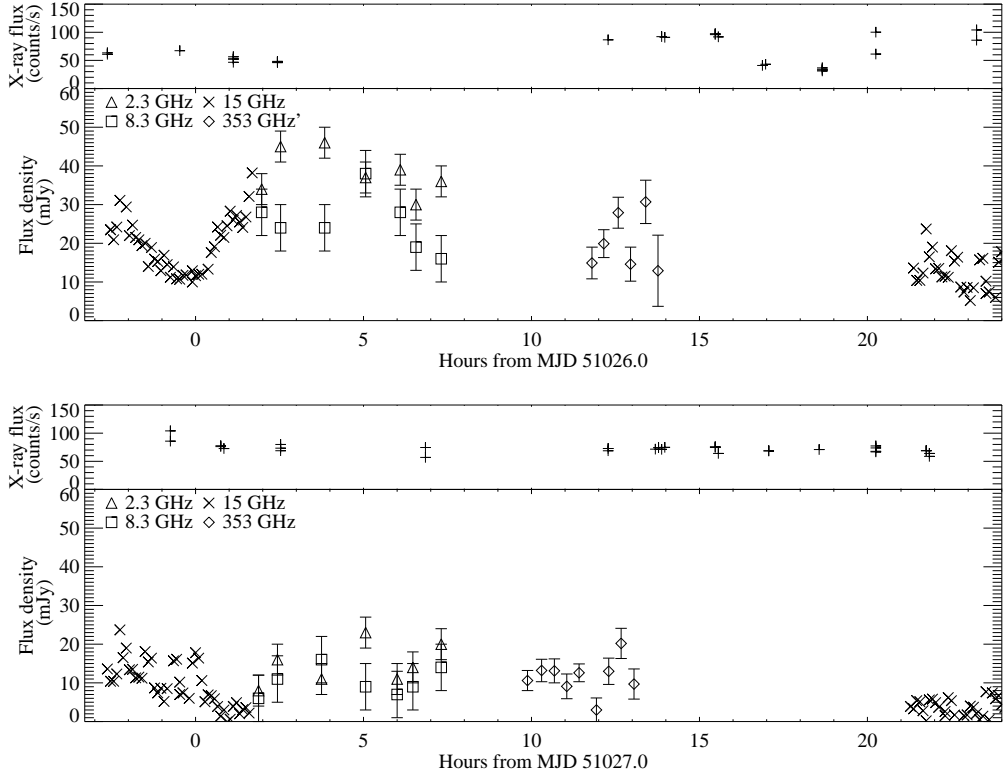, angle=90, width=6in}
\caption{Detailed photometry at radio, sub-mm and X-ray wavelengths.
The upper panel in both plots shows the X-ray flux from the {\it RXTE}
ASM, whereas the lower panel shows the radio and sub-mm data.  Radio
data at 2.25 and 8.3 GHz is from the GBI, 15 GHz is from the RT and
350 GHz is from the JCMT.}
\label{1026,1027_photometry}
\end{figure*}

\begin{table}
\caption{Photometric flux over the two epochs.  Data given is the flux
range, the average photometric error, a weighted mean, a weighted
error, a projected flux and error (see text).}
\begin{tabular}{lllllll}
\hline
$\nu$ & Range & Error & $\bar{S}$ & $\sigma_{\rm s}$ & $\bar{S}_{\rm p}$ & $\sigma_{\rm S_{\rm p}}$\\
(GHz) & (mJy) & (mJy) & \multicolumn{4}{c}{(mJy)} \\
\hline
\multicolumn{7}{c}{MJD 51026} \\
2.25	& 30--46	& 4	& 38.1	& 2.2 	& 26.1	& 1.9	\\
8.3	& 16--38	& 6	& 25.3	& 2.7 	& 18.7	& 2.3	\\
15	& 12--32	& 4	& 16.1	& 2.1 	& 12.4	& 1.6	\\
350	& 13--31	& 5	& 20.5	& 2.8 	& 20.8	& 2.8	\\
\hline
\multicolumn{7}{c}{MJD 51027} \\
2.25	& 8--23		& 4	& 14.7	& 2.0 	& 11.7	& 1.5	\\
8.3	& 6--16		& 6	& 10.3	& 1.3 	& 8.2	& 0.9	\\
15	& 2--14		& 4	& 10.2	& 1.4 	& 5.8	& 0.8	\\
350	& 3--13		& 3	& 11.5	& 1.3 	& 11.2	& 1.3	\\
\hline
\end{tabular}
\label{weighted_average}
\end{table}

\section{Interpretation}

\begin{figure*}
\epsfig{file=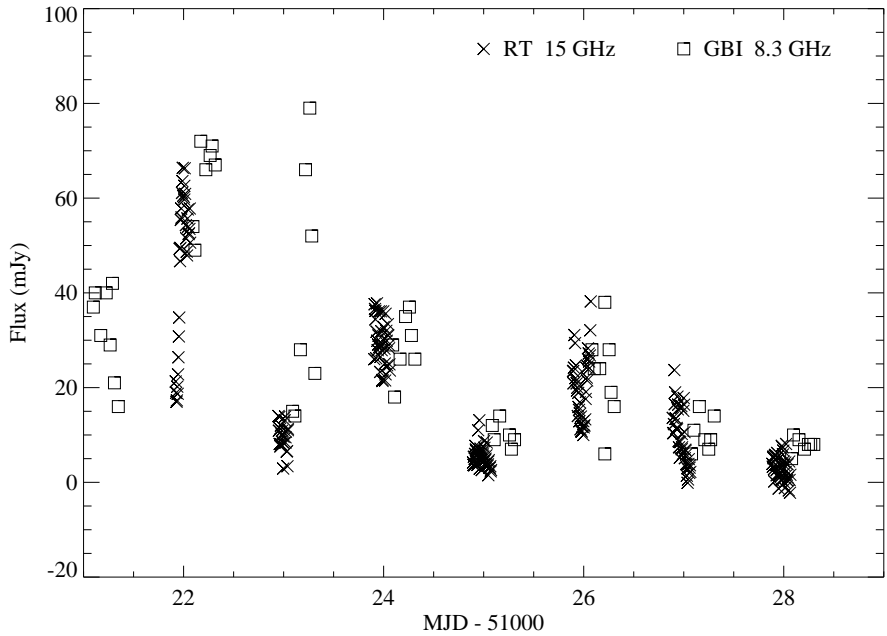, width=5in}
\caption{Detailed photometry at 8.3 and 15 GHz.  One can see that
apart from MJD 51023, behaviour at the two frequencies is similar.
Data from MJD 51022--51025 are rather variable, however the source
appears to be quietening down during the time of the SCUBA
observations at MJD 51026 and 51027.}
\label{RT_GBI_decay}
\end{figure*}

In all radio, sub-mm and X-ray bands our data show GRS 1915+105 to be
variable.  At all wavelengths, there is a general trend to lower
fluxes and smaller variability over the timescale of our observations.
Fig.\ \ref{RT_GBI_decay} shows the photometry from the GBI at 8 GHz
and the RT at 15 GHz for 7 d around the time of our SCUBA
observations.  The flux is in general decaying, with some fluctuations
around days 51023 and 51025.  The decay becomes steadier around the
time of the sub-mm observations on days 51026 and 51027.

We have modelled the decay of the last three days shown in Fig.\
\ref{RT_GBI_decay} at all three radio wavelengths with a linear fit.
This provides a plausible estimate to the cm-wave fluxes when
interpolated to the times when the sub-mm data were taken, providing
there were no rare sharp outbursts.  Averaged projected fluxes using
this method are given in Table \ref{weighted_average}.

\begin{figure*}
\epsfig{file=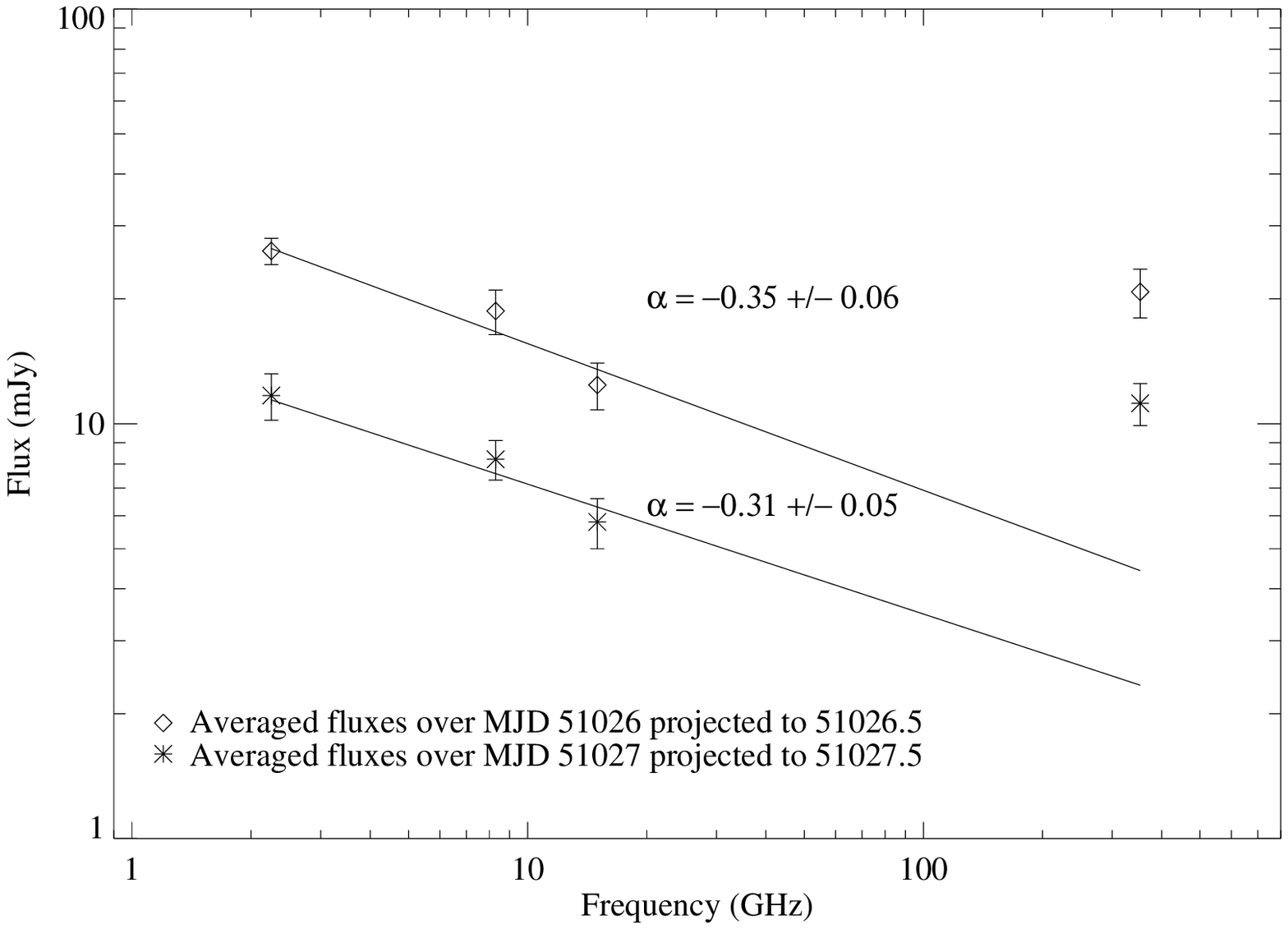, width=5in}
\caption{Two spectra using the averaged projected fluxes given in
Table \ref{weighted_average}. Diamond-shapes are the average projected flux
over MJD 51026 and the star-shapes are the average projected flux over
MJD 51027.  Note that the spectral slope over all the data is similar
and there appears to be a sub-mm excess during each epoch.}
\label{proj_spectra}
\end{figure*}

\subsection{Evidence for a high frequency excess}

The main difficulty in interpretation of these data is the validity of
extrapolating a variable source using a simple constant decay.  If
data at all radio frequencies follows the simple decay law, and no
additional flares were produced at the time of the SCUBA observations,
then the spectra over the two epochs in Fig.\ \ref{proj_spectra} shows
a clear sub-mm excess.  While these last two assumptions may be over
simplifying the reality of GRS 1915+105 (e.g.\ fig.\ 4 in PF97 shows a
large short-lived flare), previous observations have shown a
high-frequency excess at infrared compared with radio observations
(Fender \& Pooley 1998; Mirabel et al.\ 1998).  Photometric
observations by these authors have shown that over a large range in
radio flux (10--100 mJy), and with multiple oscillations, the radio
spectrum at a time when the flux is low ($\simeq 20$ mJy) is very
similar to that in this paper (e.g.\ fig.\ 1 in Mirabel et al.\ 1998),
and during times when the radio flux at 8 GHz is around 15 mJy, the
2.2 $\umu$m flux is 35-40 mJy.  

\subsection{Origin of the sub-mm excess}

The flat spectrum ($\equiv$ high-frequency excess) presented here is
seen in other X-ray binaries where there exists significant emission
at the mm and sub-mm regimes.  Both Cyg X-1 (Fender et al.\ 2000) and
Cyg X-3 (Fender et al.\ 1995; Ogley et al.\ 1998) have spectra with
indices $\alpha \simeq 0$, deviating from the expected optically thin,
$-1 \leq \alpha \leq -0.5$, steep spectrum observed during an outburst
(where spectral index $\alpha$ as $S_{\nu} \propto \nu^{\alpha}$).

As discussed by other authors, a sub-mm excess and approximately flat
synchrotron spectrum are likely to originate in a partially
self-absorbed jet (Blandford \& K\"onigl 1979; Hjellming \& Johnston
1988; Fender et al. 2000). In such a scenario higher frequencies probe
smaller emitting regions close to the base of the outflow.  Indeed,
for an ideal Blandford \& Konigl conical jet characteristic size
$\propto \nu^{-1}$, and so the regions responsible for the emission
observed at 350 GHz may be only 5\% as far downstream as those
observed at 15 GHz.  Dhawan, Mirabel \& Rodr\'{\i}guez (1998) present a
VLBI image of the core of the system in which the 15 GHz emission
arises on an angular scale of a few milliarcsec. At 350 GHz, the
emission may therefore be coming from angular scales as small as 200
microarcsec -- at 11 kpc (Fender et al. 1999) this corresponds to a
physical scale of about an astronomical unit. For a velocity of $\geq
0.9c$ (Mirabel \& Rodr\'{\i}guez 1984; Fender et al. 1999) this distance
would be traversed within a few tens of seconds. Pooley \& Fender
(1997), Mirabel et al. (1998) and Fender \& Pooley (1998) report
frequency-dependent delays in the synchrotron emission, in the sense
that higher frequecies rise and peak earlier.  Mirabel et al. (1998)
fit the delays to the form $\Delta t \propto v^{-2/3}$, although in
the self-similar regime of an ideal Blandford and Konigl jet this
should be $\Delta t \propto v^{-1}$. Either way, the for an observed
delay of tens of minutes at 15 GHz, we would also expect from this
scaling that the sub-mm emission arises less than a minute downstream
in the flow from the base of the jet. So scaling with respect to both
direct imaging and frequency-dependent delays show that sub-mm
observations are ideal for studying emission near the base of the jet
(and do not suffer the instellar extinction which can be so dramatic
for $\lambda \leq 5\mu$m).

Fig. 3 shows the radio--(sub)mm spectrum observed from GRS 1915+105 on
the two consecutive days of observation. The similar fractional
decrease in all bands strongly supports our interpretation of the
sub-mm emission as being a high-frequency extension of the synchrotron
spectrum observed at cm wavelengths. As noted above, the flow time of
electrons from the disc/black hole to the sub-mm-emitting region is
likely to be much shorter than the interval between the observations,
and so we are likely to be observing two different populations of
electrons on the two days of observations. The similarity of the
spectra is therefore an indication of the steadiness of the disc:jet
coupling and of the outflow itself. The decrease in mean flux density
between the two days may reflect a small decrease in accretion rate.
If there had {\em not} been a renewal of the electron population over
the 24 hr between the observation, this would require the magnetic
field at the site of the 350 GHz emission to be $\leq 10$ G in order
that synchrotron losses had not become significant.

We do not have a good explanation at present for the apparent minimum
in the spectrum between 15 -- 350 GHz, but given the non-simultaneity
of the data and notoriously unpredictable nature of the source, we
will not discuss it further here. In addition we note that radio--mm
observations of GRS 1915+105 presented in Fender \& Pooley (2000)
constitute evidence that the radio--infrared spectrum can steepen from
$\alpha \sim 0$ to more positive values.

\section{Conclusions}

We have obtained the first sub-mm photometry of GRS 1915+105 at
350 GHz (850-$\umu$m), together with radio data on two separate epochs
during a time when the source was mildly active.  At all frequencies a
significant amount of variability was observed, with a roughly flat
spectral index, supporting the hypothesis that emission from radio to
infrared wavelengths is dominated by synchrotron emission. Scaling
with respect to both direct VLBI imaging and radio--mm--infrared time
delays indicate that future sub-mm observations have the potential to
probe very close to the base of the jet.

\section{Acknowledgments}

The authors are very grateful for the `wizardry' of scheduling by
Graeme Watt at the JCMT, and for assistance by Richard Prestage, Rob
J.\ Ivison, Wayne Holland and Tim Jenness in co-ordinating and
reducing the data.

The Green Bank Interferometer is a facility of the National Science
Foundation operated by the NRAO in support of NASA High Energy
Astrophysics programmes.  Radio astronomy at the Naval Research
Laboratory is supported by the Office of Naval Research.  The Ryle
Telescope is supported by PPARC.  The James Clerk Maxwell Telescope is
operated by The Joint Astronomy Centre on behalf of the Particle
Physics and Astronomy Research Council of the United Kingdom, the
Netherlands Organisation for Scientific Research and the National
Research Council of Canada.

\end{document}